\documentclass[a4paper,twoside,12pt,english]{article}

\usepackage[T1]{fontenc}
\usepackage[latin1]{inputenc}
\usepackage{babel}
\usepackage{amsmath}
\usepackage{latexsym}
\usepackage{graphicx}

\def\beq{\begin{equation}}
\def\eeq{\end{equation}}
\def\bitem{\begin{itemize}}
\def\eitem{\end{itemize}}
\def\bear{\begin{array}}
\def\ear{\end{array}}

\def\lamda{\lambda}
\def\munu{{\mu\nu}}

\def\T{\hat{T}}
\def\d{\partial}

\def\dn{\d_{\nu}}
\def\dr{\d_{\rho}}
\def\q{\dot{q}}

\def\m{\hat{m}}
\def\j{\hat{j}}

\begin{document}


\begin{titlepage} \vspace{0.2in} 

\begin{center} {\LARGE \bf 

Dynamics of Matter in a Compactified Kaluza-Klein Model}\\ 
\vspace*{1cm}
{\bf Valentino Lacquaniti $^{1,2,3}$ and Giovanni Montani $^{1,4,5}$}\\
\vspace*{1cm}
$^1$ ICRA---International Center for Relativistic Astrophysics, 
Physics Department (G9),
University  of Rome, "`La Sapienza"', 
Piazzale Aldo Moro 5, 00185 Rome, Italy.\\
e-mail: valentino.lacquaniti@icra.it, montani@icra.it\\ 

$^2$  Physics Department  "`E.Amaldi "`,  University of Rome, "`Roma Tre"', 
Via della Vasca Navale 84, I-00146, Rome , Italy \\ e-mail: lacquaniti@fis.uniroma3.it \\

$^3$ LAPTH -9, Chemin de Bellevue BP 110 74941 Annecy Le Vieux Cedex, France \\

$^4$ ENEA- C. R. Frascati ( Department F. P. N. ), via E.Fermi 45, I-00044, Frascati, Rome, Italy \\
$^5$ ICRANET - C.C. Pescara, Piazzale della Repubblica 10, I-65100, Pescara, Italy \\

\vspace*{1.8cm}

{\bf   Abstract  \\ } \end{center} \indent
A longstanding problem in Kaluza-Klein models is the description of  matter dynamics. Within the 5D model,  the dimensional reduction of the geodesic motion for a 5D free test particle  formally restores electrodynamics, but the  reduced 4D particle shows a charge-mass ratio that is upper bounded, such that it cannot fit to any kind of elementary particle. At the same time, from the quantum dynamics viewpoint, there is the problem of the huge massive modes generation. We present a criticism against the 5D geodesic approach and face the hypothesis that in Kaluza-Klein space the geodesic motion does not deal with the real dynamics of  test particle. We propose a new approach: starting from the conservation equation for the 5D matter tensor, within the Papapetrou multipole expansion, we  prove that the 5D dynamical equation differs from the 5D geodesic one. Our new equation provides right coupling terms without bounding and in such a scheme the tower of massive modes is removed.

\end{titlepage}


\section*{Introduction}
In this paper we focus on the 5D Kaluza-Klein ( KK ) model, with a small compactified extra dimension.
The original KK  model  was first presented   by
Kaluza (\cite{kaluza}) and Klein (\cite{klein1}, \cite{klein2})
and successively improved by Jordan (\cite{jordan}) and Thirry (\cite{thirry}). It was a 5D model which unified gravity and electromagnetism; it can be regarded as a low-energy effective theory, in which the 5D Poincar\'e symmetry is broken, as well as the 5D Equivalence Principle ( PE ), and U(1) gauge symmetry appears. After this, KK model had a full development with the geometrization of non-abelian Yang-Mills fields (\cite{cianfranimarrocco}, \cite{montanikk}, \cite{modernkk}, \cite{cremmer}), and the idea of multidimensions was adopted in String and Supergravity theories (\cite{duff}, \cite{douglas}).
Despite this success, KK models did not have the same impact  in their  application on cosmology or quantum particles theories.
Actually, while KK models are able to reproduce correctly the dynamics of Yang-Mills fields in vacuum, they lead to  unphysical outcomes when matter is taken into account,  especially as far as elementary particles are concerned.
A longstanding problem, in the  5D model, is the correct evaluation of the charge-mass ratio for a test particle (\cite{vandongen}, \cite{overduinwesson}): the classical motion of a 5D test particle is reduced to the motion of a 4D particle interacting with electromagnetic field, but
  the charge-mass ratio results upper bounded in such a way that  any known elementary particles satisfies it. At the same time, when we consider the 5D Klein Gordon dynamics, the requirement of the compactification at a small scale leads to a tower of unphysical huge massive modes (\cite{vandongen}, \cite{overduinwesson}).
  Nowadays, there are some  alternative multidimensional models, which solve the matter coupling problem, like  induced-matter theory by Wesson(\cite{wesson}, \cite{overduinwesson}) or Randall-Sundrum models (\cite{randall1}, \cite{randall2}), which are well grounded and in full development . In these models, original hypotheses of KK are in some way changed or relaxed. What we want to do now is to pursue the original KK model; in our opinion the problem does not relies in the compactification hypothesis but in the approach usually considered to describe the dynamics of test particles, i.e. the geodesic approach.
Our idea is  the following: due to the violation of 5D PE, in the 5D model the geodesic equation only represents the path of minimum distance between two fixed points, but does not deal with the motion of a test particle. The right equation of motion has to be looked for in the 5D  conservation equation for the energy-momentum tensor, via the Papapetrou multipole expansion (\cite{pap}). This is indeed a  difference with respect to standard General Relativity (GR), where we are used to identify the  geodesic equation with  the single-pole approximated equation  we get from the energy-momentum conservation  equation. 
In this paper we will re-analyze  the geodesic approach in 5D and we will focus our criticism on the statement underlying  such an approach. We will discuss how the  idea itself of a 5D test particle  is not well defined in this scheme. Indeed we propose a new approach to solve such a problem,  based on Papapetrou expansion (\cite{pap}). We will be able, within the original compactified KK framework, to give a new equation of motion, with right coupling factors and no bound arising, where the tower of massive modes is removed. 

Paper is organized as follows: in sec. (\ref{sec:kkmodel}) we focus on fundamental statement of KK compactified model. In sec. (\ref{cd}) we examine the classical motion of a test particle; actually sec. (\ref{geodapp}) is just a review of the geodesic approach where the $q/m$ problem is found, but we add  further analysis on this topic in sec. (\ref{sec:hamiltonian}) by considering the dynamics from the Hamiltonian point of view. Section  (\ref{papapp}) is the core of the paper; we propose our new approach and discuss the new equation of motion provided by the multipole expansion of the conservation law for the 5D matter tensor. In sec. (\ref{kgd}) we consider the dynamics in the Klein-Gordon ( KG ) scheme; in sec (\ref{kgg}) we turn back to the geodesic approach, but we develop the corresponding KG equation and show how huge massive modes are linked to the $q/m$ problem. In sec (\ref{kgp}), we study the KG equation that arises from our new approach  and we get the important result that the tower of massive modes is removed. In sec. (\ref{conc}) conclusions and perspectives follow. 

\section{Kaluza-Klein 5D model}
\label{sec:kkmodel}
In the 5D  KK  model we generalize the geometrical framework of  GR  by adding a space-like dimension. We introduce the 5D metrics $J_{AB}$\footnote{we use the convention $A=0,1,2,3,5=\mu,5$}, the 5D curvature scalar ${}^5\!R$, the 5D Einstein-Hilbert Action $S_5$ and the 5D Newton constant $G_{(5)}$:
\beq
ds^2_{5}=J_{AB}dx^A dx^B 
\quad\quad
S_5=-\frac{1}{16\pi G_{(5)}}\int\!\!d^5\!x \sqrt{J}\,\, {}^5\!R    
\label{s5}
\eeq 
We are interested in the compactified KK model , based on  the following hypotheses: ( KK hypotheses ):
\bitem
\item (\textit{Compactification hyp.}) Our manifold  is the direct product $\mathcal{M}^4\otimes S^1$, where $\mathcal{M}^4$ is the ordinary space-time and $S^1$ is a space-like loop. To make the extra dimension unobservable, we assume that its size is below our observational bound: $L_5=\int \!\sqrt{J_{55}}\,dx^5\,<10^{-17} cm$.
\item (\textit {Cylindricity hyp.}) All components of the metric tensor do not depend on the extra coordinate $x^5$. Therefore, we have an invariant Killing vector with components $(0,0,0,0,1)$. We can figure this statement as the zero-order approximation of the Fourier-expansion on the extra coordinate.
\item (\textit{Scalar hyp.}) The $J_{55}$ component of the metrics is a scalar.
\eitem
Due to last two hypotheses, the model is no more invariant with respect to a generic diffeomorphism. It is actually invariant with respect to the so-called KK transformations:
\beq
\left\{
\begin{array}{l}
x^5=x^{5'}+ek\Psi(x^{\mu'}) \\
\\
x^{\mu}=x^{\mu}(x^{\nu'})
\end{array}
\right.
\label{kktrasf}
\eeq 
Here $\Psi$ is an arbitrary scalar function and $ek$ an appropriate dimensional factor to be fixed.
In agreement with the above class of transformations, the metrics and the Action admit the following reduction:
\beq
J_{AB}{}\Rightarrow{}\left(
\begin{array}{ccc}
g_{\munu}-\phi^2(ek)^2A_{\mu}A_{\nu} & -\phi^2(ek)A_{\mu} \\
                 &            \\
-\phi^2(ek)A_{\mu}     & -\phi^2
\end{array}
\right)
\label{splittingmetrics}
\eeq
\beq
S_5=-\frac{1}{16\pi {G_5}'} \int\!d^4\!x \sqrt{-g}\, (\phi R-2\nabla_{\mu}\d_{\nu}\phi+\frac{1}{4}(ek)^2\phi^3F_{\munu}F^{\munu} )
\label{kkaction}
\eeq
With respect to KK transformations, $g_{\munu}$ is a 4D tensor, then we address to it the gravitational field; $R$ is the associated 4D curvature scalar. The field $A_{\mu}$ is a U(1) gauge field ( we say the electromagnetic one ) and $F_{\munu}$ is the Faraday tensor. We also have the additional scalar field $\phi$, that is the scale factor of the extra dimension. Finally, we label with $\nabla_{\mu}$ the 4D covariant derivative and we have
 ${{G_5}'}^{-1}={G_5}^{-1}\int\!dx^5$.
 In the original KK model there was no scalar field, because from the beginning there was the condition $J_{55}=-\phi^2=-1$. In such a way the Einstein-Maxwell theory is  restored if we take  
 \beq
G={ G_5}' \quad \quad
\frac{4G}{c^4}=(ek)^2
\label{ek}
\eeq
In modern researches the scalar field is allowed, although  its presence is still matter of debate,  and, as pointed out by some authors, it  can be viewed as a massless boson of zero spin
(\cite{cianfranimarrocco}, \cite{modernkk}, \cite{witten}).  Eventually, the Einstein-Maxwell theory could be restored in the contest of varying fundamental constant models
( \cite{dirac}, \cite{barrow} ),  by taking
\beq
G={\phi^{-1} G_5}' \quad \quad
\frac{4G}{c^4}=(ek)^2\phi^2
\label{ekvaria}
\eeq
In such a case, we have to require a behavior of $\phi$ that would fit to present observed values and their limits and allows us to neglect its derivatives in the Action.
In this paper we focus only on the matter problem and we do not enter in this topic;  we will discuss either the case $\phi=1$ and the case with a varying  $\phi$.
\\
\\
A last point should be stressed in view of  following analysis.
We note that, due to the law of transformation  of metrics components  ( that does not admit non-linear transformations as far as $x^5$ is concerned ),  there is no way to find a local transformations which could bring us to a Minkowsky 5D space ( except for the case $A_{\mu}=0$, $\phi=1$, which is not interesting to us ). This means that the 5D  PE  is broken (  \cite{landau} ). The 4D PE is however still safe, because we can always find a 4D Mynkowskian space, but the breaking of the 5D one is a feature which deserves  our attention, and we will  turn back on this point.
\section{Classical Dynamics}
\label{cd}
\subsection{Geodesic Approach}
\label{geodapp}
The standard approach to the free test particle  in GR relies in the equivalence between the motion of the particle and the geodesic trajectory on the manifold:  if we associate the geodesic to the Action $S_g=\int ds$, then the dynamics of the particle is governed by the Action $S=-mS_g$ ( adopting $c=1$ ) and the fact that the mass parameter is constant and does not appear in the motion equation is consistent with the PE.
Assuming that such an equivalence still holds in 5D, we face the problem of the free test particle motion in KK via the following Action:
\beq
S_5=-\m\int ds_5 \quad\quad ds_5^2=ds^2-\phi^2 (ekA_{\mu} dx^{\mu}+d x^5)^2
\label{act5}
\eeq
where the mass parameter $\m$ is taken as a constant; in principle we have to determine whether $\m$ represents the 4D mass or a new 5D mass but at this step $\m$ does not enter in the motion equation, therefore we now don't care  this problem.  The splitting of the line element is obtained in agreement with the reduction of the metrics (\ref{splittingmetrics}). 
After  the variational procedure, followed by a reparametrization of the equations in terms of the element $ds$, we get the following set of Eulero-Lagrange equations:
\beq
\left\{
\bear{l}
\frac{d}{ds}w_5=0 \\
\\
\frac{D}{Ds}u^{\mu}=F^{\munu}u_{\nu}\left(\frac{ekw_5}{\sqrt{1+\frac{w_5^2}{\phi^2}}}\right)+\frac{1}{\phi^3}(u^{\mu}u^{\nu}-g^{\munu})\dn{\phi}\left(\frac{w_5^2}{1+\frac{w_5^2}{\phi^2}}\right)
\ear
\right.
\label{geodmotion}
\eeq
Here  $u^{\mu}=\frac{dx^{\mu}}{ds}$ is the usual 4D velocity ,$\frac{D}{Ds}$ is the 4D covariant derivative along the path and $w_5=J_{A5}w^{A}$ , while $w^{A}=\frac{dx^A}{ds_5}$.
The first equation defines a constant of motion, in agreement with the existence of a Killing vector.  Explicitly it reads as follows:
\beq
w_5=-\phi^2(ekA_{\mu}w^{\mu}+w^5)
\label{w5}
\eeq
A direct calculus would show moreover  that $w_5$ is a scalar with respect to KK transformations \footnote{ Actually, this is a generic feature of KK model: for a given vector in covariant representation its fifth component is always a scalar }. 
The second equation gives the dynamics of the reduced interacting particle. Clearly our system is not purely electrodynamics due to presence of $\phi$. Anyway, we can restore the electrodynamics coupling in terms of the observed charge $q$ and mass $m$ of the particle,  by setting
\beq
\frac{q}{m}=ek\frac{w_5}{\sqrt{1+\frac{w_5^2}{\phi^2}}}
\label{qm}
\eeq
While this definition sounds perfectly reasonable, as well as all steps of this procedure, it leads to an unphysical feature.
Indeed, recalling eq.(\ref{ekvaria}) we now have:
\beq
\frac{q^2}{4Gm^2}=\frac{w_5^2}{\phi^2}\frac{1}{1+\frac{w_5^2}{\phi^2}}
\eeq
This means that, not depending on the value of $\phi$ , the right member is always upper bounded and we have:
\beq
\frac{q^2}{4Gm^2}< 1
\eeq
Every known elementary particle does not satisfy this 
bound. If we consider the electron, for instance, we have $\frac{e^2}{4Gm_e^2}\sim (10^{43})$. This feature still holds even if we take $\phi=1$ from the beginning and consider a purely electrodynamics model. Eventually, the presence of $\phi$ gives the additional feature that the $q/m$ ratio is not conserved. The $q/m$ puzzle ( \cite{vandongen}, \cite{overduinwesson} ) is a longstanding problem that affects KK models, and it is strictly linked, as we will see in the following ( sec. \ref{kgg} ), to the problem of the generation of huge massive modes beyond the Planck scale. Hence, from this viewpoint, the KK model seems not to be able to describe the coupling with matter. 


\subsubsection{Hamiltonian Approach}
\label{sec:hamiltonian}
So far, we have seen a review of the well known geodesic approach. Now, let us consider 
 the Hamiltonian reformulation of the dynamics which will give us a nice insight .
Indeed, the analysis of conjugate momenta and dispersion relation will allow us  to isolate the contribution  of charge with respect to the contribution of mass in the formula (\ref{qm}). The calculus of the Hamiltonian concerning the Action (\ref{act5}) is quite tedious and requires the ADM reformulation of the model (\cite{adm}, \cite{kolb}, \cite{thiemann}, \cite{lacquaniti}).  
We    sketch the procedure in the next paragraph  
 and analyze  thereafter the outcome . 
\paragraph{\emph{Hamiltonian formulation: }}
Let us consider the Action (\ref{act5}):
$$
S_5=-\m\int ds_5 \quad\quad ds_5^2=ds^2-\phi^2 (ekA_{\mu} dx^{\mu}+d x^5)^2
$$
To get the Lagrangian we  define a time-like coordinate $q^0$ and  perform the following reparametrization:
\beq
S_5=-\m\int\!dq^0\frac{ds_5}{dq^0} \quad\quad\rightarrow L=-\m\frac{ds_5}{dq^0}
\label{lagg}
\eeq
  Following formulas provide the ADM  (\cite{adm}, \cite{kolb}, \cite{thiemann}, \cite{lacquaniti}) splitting of our variables: 
\begin{eqnarray}
A_{\mu}&=&(A_0,A_i)\\
dx^{\mu}&=&  (dq^0,-N^idq^0-dq^i) \\
dx_{\mu}&=&  (dq_0,dq_i ) 
\end{eqnarray}
Components $dq_i$ and $A_i$ represent the spatial projection of $dx_{\mu}$ and $A_{\mu}$ and we have  $dq_i=h_{ij}dq^j$, $dq^0=\frac{1}{N^2}(dq_0-dq_iN^i)$, where $N$, $N^i$, $h_{ij}$ are respectively  the \textit{Lapse} function, the \textit{Shift} functions and the 3D induced spatial metrics. We define the following set of Lagrangian coordinates and velocities:
\begin{eqnarray}
q^{\mu}&=&(q^0,q^i)\quad q^5=-x^5 \nonumber \\
\q^i&=&\frac{dq^i}{dq^0} \quad \q^5=\frac{dq^5}{dq^0} 
\label{lagset}
\end{eqnarray}
By substituting these formulas into the line element $ds_5^2$ , we  calculate the Lagrangian (\ref{lagg}) and we obtain:
\beq
L=-\m[N^2-h_{ij}(\q^i\q^j)-\phi^2(-\q^5+ekA_0-ekA_jN^j -ekA_j\q^j)^2]^{\frac{1}{2}}
\eeq
From the above expression we can  calculate conjugate momenta $P_i,  P_5$ and perform the Legendre transformation. In a synchronous frame, where \textit{Shift} function vanish and \textit{N} reduces to unity we have:
\beq
(H-ekP_5A_0)^2= h^{ij}\Pi_i\Pi_j+\left(\m^2+\frac{P_5^2}{\phi^2}\right)
\label{disprel}
\eeq
where $\Pi_i=P_i-ekP_5A_i$, $H$ is the Hamiltonian and we have $
\Pi_i=\m w_i, P_5=\m w_5
$,
being $w^A=\frac{dx^A}{ds_5}$. If we identify $H$ with the time component of the momentum we can write a 4D covariant relation which holds in every frame:
\beq
g^{\munu}\Pi_{\mu}\Pi_{\nu}=\m^2+\frac{P_5^2}{\phi^2},
\label{gianni}
\eeq
where $\Pi_0=P_0-ekP_5A_0$.
Finally, we can rebuild a 5D dispersion relation: we define $P^A=\m w^A$. It yields:
\begin{eqnarray}
P^A&=&(\Pi^{\mu}, P^5) \\
P_{A}&=&(\Pi_{\mu}+P_5A_{\mu}, P_5)\\
\Pi^{\mu}&=&g^{\munu}\Pi_{\nu}\quad\quad P_5=-\phi^{2}(A_{\mu}\Pi^{\mu}+P^5).
\end{eqnarray}
With such a definition we can reproduce the relation (\ref{gianni}) by the following 5D dispersion relation :
\beq
P_AP^{A}=\m^2.
\label{5ddisprel}
\eeq
\\

For sake of simplicity, we just restrict  to the synchronous frame where the relation (\ref{disprel}) holds and we have
\beq
P_i=\Pi_i+ekA_iP_5.
\label{minsub}
\eeq
It is worth stressing that 
the momenta $P_i,P_5$    are  proportional to $w_i,w_5$ through the parameter $\m$ and the momentum $P_5$ is scalar and conserved, being conjugated to a cyclic coordinate. We   identify the relation (\ref{disprel}) with the dispersion relation of a 4D  test particle with charge $q$ and mass $m$ in an external electromagnetic 
  field, where we define
 \begin{eqnarray}
 \label{43}
 q&=& ekP_5 \\
 \label{44}
 m^2&=& \left(\frac{P_5^2}{\phi^2} +\m^2\right) 
 \end{eqnarray}
 At the same time, we can identify relation (\ref{minsub} ) with the minimal substitution we are used to do in electrodynamics.  Now, we observe that via the formula $P_5^2=\m^2 w_5^2$ we can write
 \beq
\frac{q^2}{m^2}=(ek)^2\frac{P_5^2}{\m^2+\frac{P^2_5}{\phi^2}}=(ek)^2\frac{w^2_5}{1+\frac{w^2_5}{\phi^2}}
\eeq
Therefore we recover the formula (\ref{qm}) previously discussed.
But, now we have a fine indication:  the charge $q$ is still conserved ( as correct in a U(1) gauge invariant model ), and the mass term $m$ only carries the factor $\frac{P_5^2}{\phi^2}$, which violates the conservation and causes  the upper bound of the ratio. Moreover, we see that the mass parameter $\m$ we put in the Action does not represents the physical rest mass of the particle. Hence, while in GR the formulation of the test particle Action  
takes for granted that the mass parameter which  appear in the Action represents  the physical rest mass, now things seem very different. In our opinion, is   the misidentification  of the mass  that causes  the upper bound on the $q/m$ ratio.

\subsection{Revised Approach: Papapetrou Expansion}
\label{papapp}
\subsubsection{Criticism versus the Geodesic Approach}
The unphysical feature provided by the geodesic approach seems to show that the KK model is not able to describe the dynamics of particle.  Our opinion is different: KK model is well able to describe particles, but the geodesic approach is not the correct method to face the problem. In such an ansatz  the geodesic equation (\ref{geodmotion}) simply represents the minimum path between two points in the 5D space, but this trajectory does not deal with the dynamics of the particle.
Indeed, when we state that it is possible to describe particles via the Action (\ref{act5}), we implicitly make three assumptions, borrowing them from GR: i) $\m$ represents the rest mass ii) $\m$ is a constant iii) is possible to consider a 5D point-like particle.
Let us look at the first two assumption: in GR they are supported by the PE; actually the physical equivalence between the geodesic Action and the particle Action is an alternative statement of the PE. But, we know that in KK the 5D PE is violated, therefore in 5D such assumptions are not well grounded. Without PE, the meaning of the parameter $\m$ is quite ambiguous and in principle the 5D mass could be not constant: at least, we have seen that it does not represents the rest mass of the reduced particle. Hence, we are legitimated to take the hypothesis that the 5D geodesic equation 
represents only the trajectory of minimum path between two given points, but does not deal with the 
right equation of motion of our test particle. Now we look at the third  assumption with the following reasoning: 
 in order to correctly define a 5D test particle we have to assume that is possible to localize it. But, if we require that the size of the fifth dimension is $L_5\sim O(10^{-17})\, cm$ or smaller ,  to have a particle localized in the extra dimension we have to assume it to have an enormous amount of energy, beyond  the scale of $Tev$ , i.e.
$
E\sim O\left (\frac{1}{L}\right)
$. 
Hence, we suppose that the particle is not localized around the extra dimension and the definition itself of a 5D test particle is ambiguous.
Now, let us examine how these assumptions work in GR:
indeed, in the standard 4D theory we have a well grounded procedure which allows us to i) face a rigorously definition of rest mass ii) prove that mass is scalar and constant iii) prove that the motion of a particle is really the geodesic one when we deal with the point-like approximation for the particle. This procedure is the multipole expansion, and is  due to Papapetrou (\cite{pap}). In such an approach we deal with the equation for a generic kind of matter described by its energy-momentum tensor $T^{\munu}$, i.e.
$
 \nabla_{\mu}T^{\munu}=0
 $.  Point-like  bodies are described by the hypothesis of  localization: they are associated  to a small world-tube and components of matter tensor are non zero only  within this world-tube. Such a localization allows us to perform a multipole expansion once we integrate our equation. The test particle is given by the zero-order approximation ( single-pole ) and it corresponds to a point like particle, while , for instance, the first order ( pole-dipole interaction ) corresponds to a spinning particle. The geodesic equation, which we obtain by the Action $S=-m\int\!ds $, coincides with the single-pole equation of Papapetrou, where the mass arises  now rigorously defined as follows:
 \beq
 m=\frac{1}{u^0}\int\!\!d^3x\,\sqrt{g}\,T^{00}
 \label{papmass}
 \eeq
 Thus,  Papapetrou approach gives the proof that in standard GR the motion is effectively the geodesic one as far as the point like particle approximation is concerned.
 Therefore, our point of view is the following. The proper procedure to be generalized from 4D to 5D is not the geodesic one : we have to consider the general dynamical equation, involving a 5D matter tensor, and only after this, if we want to analyze the motion of a test particle, address the procedure of multipole expansion , which is the only one able to correctly define  the rest mass of the particle.
 \subsubsection{Multipole Expansion}
 In this approach we define a generic, symmetric, 5D matter tensor $T^{AB}$; its dynamics is given by the following equations: 
 \beq
 \left\{
 \bear{l}
 D_A T^{AB}=0\\
\d_5T^{AB}=0\\
\ear
\right.
\label{eqvera}
\eeq
where $D_A$ is the 5D covariant derivative. The first equation is the generalization of the conservation law, while the second equation is given for  consistency with the field dynamics and the cylindricity hypothesis. In order to simplify the identification of matter tensor components with corresponding 4D objects, we will  write our equations in terms of following components:
\begin{eqnarray}
&& T^{\munu} \rightarrow 4D\; tensor\nonumber \\
&& T_5^{\mu} \rightarrow 4D\; vector \nonumber \\
&& T_{55} \rightarrow \;scalar \nonumber 
\end{eqnarray}
The KK reduction of the first equation of (\ref{eqvera}), after some algebraic rearrangement,  provides the following set : 
\begin{eqnarray}
\label{eq5x}
&5)& \rightarrow \nabla_{\mu}\left(\phi T_5^{\mu}\right)=0 \\
\label{eqmux}
&\mu)&\rightarrow \nabla_{\rho}T^{\mu\rho}=-\left(\frac{\dr\phi}{\phi}\right)T^{\mu\rho}-g^{\mu\rho}\left(\frac{\dr\phi}{\phi^3}\right)T_{55}+ekF^{\mu}_{\,\,\rho}T_5^{\rho}
\end{eqnarray}
From the first of the above equations we recognize a conserved vectorial current, i.e.
\beq
j^{\mu}=ek\phi T_5^{\mu}
\label{defj}
\eeq
Therefore we rewrite:
\begin{eqnarray}
\label{eq5bis}
&5)& \rightarrow \nabla_{\mu}j^{\mu}=0 \\
\label{eqstbis}
&\mu)&\rightarrow \nabla_{\rho}\left(\phi T^{\mu\rho}\right)=-g^{\mu\rho}\left(\frac{\dr\phi}{\phi^2}\right)T_{55}+F^{\mu\rho}j_{\rho} 
\end{eqnarray}
where  $j_{\mu}=g_{\munu}j^{\nu}$.
It is worth noting  that, if we take $\phi=1$ from the beginning, our set reduces to
\begin{eqnarray}
\label{eq5fi1}
&5)& \rightarrow \nabla_{\mu}j^{\mu}=0 \\
\label{eqsf1}
&\mu)&\rightarrow \nabla_{\rho}T^{\mu\rho}=F^{\mu\rho}j_{\rho} 
\end{eqnarray}
Therefore, if we identify $T^{\munu}$ with the matter tensor of the reduced 4D matter,  we are able to take into account the coupling of matter with the conserved current associated to a U(1) gauge invariant field, namely the electromagnetic field.
 Any  hypothesis hasn't  been made on the kind of the 4D matter and the current is only defined  in terms of components of the 5D matter tensor, without any connection with the kinematics of the matter. So, at this step, no definition of mass has been employed and there is not any kind of restriction on the value of the current.
 
  Now, in order to describe in a consistent way the properties of a test particle and neglect the back-reaction, we have to adopt a localized matter tensor and follow the Papapetrou procedure.
 
 \paragraph{\emph{Papapetrou procedure: }}
Let us start from  equations (\ref{eq5bis}): we rewrite it via the identity $\nabla_{\mu}V^{\mu}=\frac{1}{\sqrt{g}}\d_{\mu}(\sqrt{g}V^{\mu})$ and moreover we perform the derivative $\d_{\mu}(x^{\nu}\sqrt{g}j^{\mu})$. Defining $\j^{\mu}=\sqrt{g}j^{\mu}$ we obtain the set
\beq
\left\{
\bear{l}
\d_{\mu}\j^{\mu}=0 \\
\d_{\mu}(x^{\nu}\j^{\mu})=\j^{\nu}\\
\ear
\right.
\eeq
Now we integrate these equations over the 3D space and use Gauss theorem. At this step we use a localization hypothesis: the current $\j^{\mu}$ is peaked on a world line $X^{\mu}$ and negligible outside: therefore we perform a Taylor expansion $x^{\mu}= X^{\mu}+\delta x^{\mu}$ and retain only the lowest order. In such an approximation our set becomes
\begin{eqnarray}
\label{c0}
\frac{d}{dx^0}\int\!\!d^3x\,\j^0=0 \\
\label{c1}
\left(\frac{dX^{\nu}}{dx^0}\right)\int\!\!d^3x\,\j^0=\int\!\!d^3x\,\j^{\nu}
\end{eqnarray}
Now , we define $q=\int\!\!d^3x\,\j^0$ and $u^{\mu}=\frac{dX^{\mu}}{ds}$, where $ds^2=g_{\munu}dX^{\mu}dX^{\nu}$. Therefore we have:
\beq
\left\{
\bear{l}
\frac{d}{ds}q=0 \\
\\
u^{\mu}q=u^0\int\!\!d^3x\,\j^{\mu}\\
\ear
\right.
\label{boh}
\eeq
Hence, we have defined the charge of the test particle. Finally we can recognize that in such point-like approximation the effective current can be written as follows:
$$
\sqrt{g}j^{\mu}=\int\!\!ds\,qu^{\mu}\delta(x^{\mu}-X^{\mu})
$$
Now we turn our attention to eq.(\ref{eqstbis}) and generalize the above procedure: at first we have to define objects that we want to localize; for the tensor component, as it happens for the vector one, the most simple choice is to consider $\phi T^{\munu}$ which is indeed the argument of the $\nabla_{\mu}$ operator. 
Then, we adopt a similar parameterization for the scalar component, thus we define now $\tilde{T}^{\munu}=T^{\munu}\phi$, $\tilde{T}_{55}= T_{55}\phi$ and apply the localization hypothesis on these objects.
Now, we use the identity $\nabla_{\rho}L^{\mu\rho}=\frac{1}{\sqrt{g}}\dr\sqrt{g}L^{\mu\rho}+\Gamma^{\mu}_{\rho\lamda}L^{\rho\lamda}$; moreover we consider the derivative of  $x^{\nu}\sqrt{g}T^{\mu\lamda}$. Defining $\T^{\munu}=\sqrt{g}\tilde{T}^{\munu}$ and $\T_{55}=\sqrt{g}\tilde{T}_{55}$, we get the set of equations:
\beq
\dr\T^{\mu\rho}+\Gamma^{\mu}_{\rho\lamda}\T^{\rho\lamda}=-g^{\mu\rho}\left(\frac{\dr\phi}{\phi^3}\right)\T_{55}+F^{\mu\rho}\j_{\rho}
\label{pap1}
\eeq 
\beq
\d_{\lamda}(x^{\nu}\T^{\mu\lamda})=\T^{\munu}-x^{\nu}\Gamma^{\mu}_{\rho\lamda}\T^{\rho\lamda}+x^{\nu}\left[-g^{\mu\rho}\left(\frac{\dr\phi}{\phi^3}\right)\T_{55}+F^{\mu\rho}\j_{\rho}
\right]
\label{pap2}
\eeq
Now we integrate over the 3D space, use the Gauss theorem and take into account the localization hypothesis. We assume that $\T^{\munu}$ and $\T_{55}$, as well as $\j_{\mu}$, are peaked on a world line $X^{\mu}$. 
Hence, the localization is performed only in the 4D space rather than in the 5D one ; this happens because the matter tensor does not depend on the fifth coordinate and it is consistent with the unobservability of the extra dimension and with the phenomenological request that we observe trajectories only in our 4D space.
Given the  displacement $\delta x^{\mu}=x^{\mu}-X^{\mu}$, we perform a Taylor expansion in powers of $\delta x^{\mu}$, with center $X^{\mu}$, as far as the Christoffel symbols and  the metric fields $\phi$ and $A^{\mu}$ (and their derivatives) are concerned. Thus, we obtain an expansion in terms of integrals like $ \int \!\!d^3x\,\delta x^{\alpha}\T^{\rho\lamda}$, $\int \!\!d^3x\,\delta x^{\alpha}\delta x^{\beta}\T^{\rho\lamda}$ and so on, while metric fields and Christoffel (and their derivatives) are carried out from the integration being estimated on $X^{\mu}$. The test particle motion is described by the lowest order of this approximation ( single pole approximation  ), where we neglect all integrals with terms of order $\delta x^{\mu}\T^{\rho\lamda}$ and greater.
Therefore, within the zero-order  approximation, equations  (\ref{pap1}, \ref{pap2}) respectively become
\begin{eqnarray}
\label{uno}
\frac{d}{dx^0}\left(\int \!\!\!d^3x \,\T^{\mu 0}\right)\!\!\!\!&+&\!\!\!\Gamma^{\mu}_{\rho\lamda}\!\int \!\!\!d^3x\,\T^{\rho\lamda}\!=-\!g^{\mu\rho}\left(\frac{\dr\phi}{\phi^3}\right)\!\!\int \!\!\!d^3x\,\T_{55}\nonumber \\
&+&\!\!\!F^{\mu\rho}\!\!\int \!\!\!d^3x\,\j_{\rho}
\end{eqnarray}
\begin{eqnarray}
\label{due}
\frac{dX^{\nu}}{dx^0}\int\!\!\!d^3x\,\T^{\mu 0}=\int\!\!\!d^3x\,\T^{\munu}
\end{eqnarray}
where now    metric fields and their derivatives are evaluated on $X^{\mu}$.
Now  we define the auxiliary tensor $M^{\munu}=u^0\int\!\!d^3x\,\T^{\munu}$, then  equations (\ref{uno}, \ref{due}) become :
\begin{eqnarray}
\label{one}
\frac{d}{ds}\left(\frac{M^{\mu 0}}{u^0}\right)+\Gamma^{\mu}_{\rho\lamda}M^{\rho\lamda}&=&-g^{\mu\rho}\left(\frac{\dr\phi}{\phi^3}\right) u^0\!\!\!\int\!\!\! d^3x\,\T_{55}\nonumber \\
&+&F^{\mu\rho}u^0\!\!\!\int\!\!\!d^3x\,\j_{\rho}
\end{eqnarray}
\beq
 M^{\munu}=\frac{u^{\nu}}{u^0}M^{\mu 0}
\label{two}
\eeq
Now, from (\ref{two}), we get immediately:
$$
M^{0\nu}=\frac{u^{\nu}}{u^0}M^{00}\,\Rightarrow M^{\munu}=u^{\nu}u^{\mu}\frac{M^{00}}{(u^0)^2}
$$
 We define the scalars $A$ and  $m$:
\begin{eqnarray}
&{}& m=\frac {M^{00}}{(u^0)^2}=\frac{1}{u^0}\int\!\!\!d^3x\,\T^{00} \\
&{}& A=u^0\int\!\!\!d^3x\,\T_{55}
\end{eqnarray}
Finally, recalling eq. (\ref{boh}),  the equation of motion (\ref{one}) reads:
\beq
\frac{D}{Ds}mu^{\mu}=-g^{\mu\rho}\left(\frac{\dr\phi}{\phi^3}\right) A+qF^{\mu\rho}u_{\rho}
\label{moto1}
\eeq
Now we require that the  identity $u_{\mu}\frac{D}{Ds}u^{\mu}=0$ holds; from the above equation we get the condition:
\beq
\frac{d}{ds}m=-\frac{A}{\phi^3}\frac{d\phi}{ds}
\label{newmass}
\eeq
Finally, by using this new condition, we can rewrite the equation of motion as follows:
\beq
m\frac{Du^{\mu}}{Ds}=\left(u^{\mu}u^{\rho}-g^{\mu\rho}\right)\frac{\d_{\rho}\phi}{\phi^3}A+qF^{\mu\rho}u_{\rho}
\label{neweq}
\eeq
Coupling factors are explicitly defined as follows:
\begin{eqnarray*}
 &{}& m=\frac{1}{u^0}\int\!\!\!d^3x\,\sqrt{g}\,\phi{T^{00}} \\
&{}& q=ek\int\!\!\!d^3x\,\sqrt{g}\,\phi{T_5^0}\ \\
&{}& A=u^0\int\!\!\!d^3x\,\sqrt{g}\, \phi{T_{55}} 
\end{eqnarray*}
Quantities $m$, $q$, $A$ result to be scalar objects.
We could obtain the same result for the motion equation and for coupling factors if we define as follows the effective matter tensor for particles :
\begin{eqnarray*}
 &{}&  \phi{\sqrt{g}T^{\mu\nu}}=\int\!\!\!ds\,m\,\delta^4(x-X)u^{\mu}u^{\nu} \\
&{}& ek \phi{\sqrt{g}T_5^{\mu}}=\int\!\!\!ds\,q\,\delta^4(x-X)u^{\mu} \\
&{}&   \phi{\sqrt{g}T_{55}}=\int\!\!\!ds\,A\,\delta^4(x-X)
\end{eqnarray*}
Hence, the effective matter tensor fits to our beginning hypothesis of localization.\\

At first we note that, if we take from the beginning $\phi=1$, then $m$ becomes a constant and our equation reduces to the Lorenz equation, once we identify $q$ with the charge and $m$ with the rest mass of the reduced particle.
Let us now compare the old equation (\ref{geodmotion}) to this new one (\ref{neweq}) in the general case. We recognize that:
\bitem
 \item they show the same dynamical structure but coupling factors are not the same .
 \item in (\ref{neweq}) we have the three factors $m$, $q$, $A$ that are defined in terms of independent degrees of freedom of the matter tensor, therefore are not correlated each to other, while in the geodesic approach $q$ and $m$ were both defined in terms of $P_5$ ( so giving the upper bound ): therefore now no bound arises.
  \item q is conserved, due to the presence of a conserved current ( gauge theory); A is not constant, but in principle there is no symmetry requiring its conservation.
   \item mass is not conserved (\ref{newmass}) and this is indeed a relevant feature of this new equation. Anyway, there is no reason to require in principle the conservation of $m$ : the 5D PE is broken and therefore mass is not necessarily a constant. However, in the scenario with $\phi=1$ we deal with a purely Einstein-Maxwell system, so the PE is restored and mass turns to be a constant.  It is worth remarking the existence of other  interesting scenarios. By setting $A=0$ we recover a constant mass and the free falling universality of the particle ( in absence of electromagnetic field ), without any constraint on the scalar field. By setting $A=\alpha m\phi^2$ , we loose a constant mass but we still have the free falling universality and an explicit form for the behavior of the mass it is easily founded by integrating the equation (\ref{newmass}):
\beq
m=m_0\left( \frac{\phi}{\phi_0}      \right)^{-\alpha}
\label{scalmass}
\eeq
\eitem
Papapetrou approach  shows that the particle motion is not the geodesic one and that the idea of a 5D test particle is misdealing, being the particle localized only in the observed 4D space,  and tells how charge and mass have to be defined correctly.
In the  next section we will  enforce this point of view showing how this new approach is able to remove the KK tower of massive modes.


\section{Klein-Gordon Dynamics}
\label{kgd}
\subsection{Standard Approach}
\label{kgg}
Within the geodesic procedure the 5D velocity of the particle must satisfy the
identity 
\beq
J^{AB}w_Aw_B=1
\eeq
If we introduce the 5D  mass parameter $\m$   and define linear momentum $P_A=\m w_A$, we gain the 5D dispersion relation (\ref{5ddisprel})
\beq
J^{AB}P_AP_B=\m^2
\label{5ddisp}
\eeq
Such a relation  provides the  dispersion relation (\ref{disprel}) we have seen in sec. (\ref{sec:hamiltonian}); actually, we could consider it as a more general formulation of eq.( \ref{disprel} ). 
We  consider the canonical quantization of this relation, as far as a complex scalar field $\zeta$ is concerned, and we get a 5D KG equation. The associated Lagrangian density reads:
\beq
\mathcal{L}=J^{AB}\d_A\zeta(\d_B\zeta)^+-\m^2\zeta\zeta^+
\eeq
Now we implement the KK reduction.  To be consistent with the unobservability request, we assume that our scalar field depends on $x^5$ only through a phase factor:
\beq
\zeta(x^{\mu},x^5)=\eta(x^{\mu})e^{iP_5x^5}
\label{phasef}
\eeq
The momentum $P_5$ is scalar and conserved. Moreover, recalling KK transformations (\ref{kktrasf}), we notice that our field transforms as follows:
\beq
\zeta=\zeta'e^{i(ekP_5)\Psi}
\eeq
Therefore, we are dealing correctly with a U(1) gauge transformation. Then, implementing the reduction of the inverse metrics $J^{AB}$, we finally can consider the reduction of the Lagrangian and this yields :
\beq
\mathcal{L}=g^{\munu}(-i\d_{\mu}-P_5ekA_{\mu})\eta[(-i\d_{\nu}-P_5ekA_{\nu})\eta]^+-\left(\m^2+\frac{P_5^2}{\phi^2}\right)\eta\eta^+
\label{51}
\eeq
We recognize a U(1) gauge invariant lagrangian where the reduced field $\eta$ acquires  a charge $ekP_5$ and a mass term $m^2=(\m^2+\frac{P_5^2}{\phi^2})$; the ratio $q/m$  fits to the result previously obtained for the motion of the test particle. Moreover, requiring the compactness of the fifth dimension, we get  the quantization of $P_5$, i.e. $P_{5(n)}={2\pi n}/{L_5}$. Thus, the discretization of $P_5$ gives rise to a tower of modes for the mass term $m$; in the simplest case, with $\phi=1$, we simply have:
$$m_{(n)}^2=\m^2+P_{5(n)}^2$$
Fixing the minimum value of $P_5$ via the elementary charge $e$, we get the extra dimension size below our observational limit, i.e.  $L_5\simeq 10^{-31}\,\,cm$. But, at the same way, this evaluation provides huge massive modes beyond the Planck scale. Therefore, within the geodesic approach, the problem of the charge-mass ratio is strictly linked to the problem of massive modes ( \cite{vandongen}, \cite{overduinwesson} ). 

 \subsection{Revised Approach}
 \label{kgp}
Let us consider now the KG dynamics that arises from the Papapetrou revised approach.
At first, we note that the new equation (\ref{neweq}) can be derived from the following Action:
\beq
S=-\int\!\!m\,d\,s+q\left(A_{\mu}dx^{\mu}+\frac{dx^5}{ek}\right)
\label{revaction}
\eeq
In agreement with equation (\ref{newmass}), the parameter $m$ has to be regarded as a variable function, not depending on $x^5$, whose derivatives are given by the following formula :
\beq
\frac{\d m}{\d x^{\mu}}=-\frac{\d_{\mu}\phi}{\phi^3}A
\eeq
 A direct calculus of Eulero-Lagrange equations would give the proof that this Action leads to eq. (\ref{neweq}). Moreover, it is worth noting that, if $\phi=1$,  being $m$ a constant in such a scenario, we have exactly the Lorenz Action plus the charge conservation law, expressed in term of a Lagrange multiplier. 
From the above Action,  we can calculate
 Lagrangian, conjugate momenta and Hamiltonian, in order to analyze the dispersion relation.
 
 \paragraph{\emph{Hamiltonian formulation:}}
Starting  from the Action (\ref{revaction}),
 we consider the ADM reformulation (\cite{adm}, \cite{kolb}, \cite{thiemann}, \cite{lacquaniti}), as well as we have done in the  section (\ref{sec:hamiltonian}): introducing the same  set of Lagrangian coordinates $q^0,q^i,q^5$ we get the following Lagrangian:
$$
L=-m[N^2-h_{ij}(\q^i\q^j)]^{\frac12}-q(-\frac{\q^5}{ek}+A_0-A_jN^j -A_j\q^j)
$$
From the above expression we  calculate the Hamiltonian $H$ and the conjugate momenta $P_i, P_5$. In a synchronous frame we have:
\beq
(H-qA_0)^2= h^{ij}\Pi_i\Pi_j+m^2,
\label{koso}
\eeq
where 
\begin{eqnarray}
P_5&=&\frac{q}{ek}\\
\Pi_i&=&P_i-qA_i=mu_i \quad\quad u^{\mu}=\frac{dx^{\mu}}{ds}.
\end{eqnarray}
Identifying the Hamiltonian with the time component of the momentum we can rewrite equation (\ref{koso}) in a 4D manifestly covariant expression:
\beq
\Pi_{\mu}\Pi^{\mu}=m^2,
\label{koso2}
\eeq
where $\Pi_0=P_0-qA_0$.
Finally, we can rebuild a 5D dispersion relation: the calculus of $P_{\mu}, P_5$ we got from the Lagrangian, uniquely defines a 5D vector $P^A$ such  that we have:
\begin{eqnarray}
P^A&=&(\Pi^{\mu}, P^5) \\
P_{A}&=&(\Pi_{\mu}+P_5A_{\mu}, P_5)\\
\Pi^{\mu}&=&g^{\munu}\Pi_{\nu}\quad\quad P_5=-\phi^{2}(A_{\mu}\Pi^{\mu}+P^5).
\end{eqnarray}
where $P_5=\frac{q}{ek}$, $\Pi^{\mu}=mu^{\mu}$.
With such a definition, relations (\ref{koso},\ref{koso2}) are provided by the following 5D dispersion relation :
\beq
P_AP^{A}=m^2-\frac{q^2}{(ek)^2\phi^2}
\label{disp5}
\eeq
\\

Therefore, looking at equations (\ref{koso},\ref{koso2}), we finally recognize the dispersion relation for a test particle with mass $m$ and charge $q$ interacting with the electromagnetic field.  Now $m$ is exactly the mass parameter we put in the Action, and at the same time is the scalar which defines the 4D dispersion relation (\ref{koso2}) and the proportionality relation between usual 4D conjugate momenta and velocities. Thus,  this result enforces our interpretation concerning the Papapetrou procedure. 
\subsubsection{Removal of the KK tower}
Moreover , it is worth noting that there is now no link  between $P_5$ and the velocity. Therefore, the quantization of charge, i.e. the quantization of $P_5$, does not affect the definition of the mass. Indeed,  let us calculate the 5D KG equation corresponding to the dispersion relation (\ref{koso}).
The formula (\ref{disp5}) , when we consider its dimensional reduction, leads to the relation (\ref{koso}), therefore we assume that it represents the 5D dispersion relation of the particle. Hence, the 5D KG equation associated to our particle has to be defined via the quantization of relation (\ref{disp5}).
When we perform the canonical quantization of eq. (\ref{disp5}), repeating the procedure of sec. (\ref{kgg}),
 in the resulting Klein-Gordon equation  we now  have the  counter term $-\frac{q^2}{(ek)^2\phi^2}$ that rules out the huge massive modes. 
 The final Lagrangian for the reduced field $\eta$  reads
 \beq
\mathcal{L}=g^{\munu}(-i\d_{\mu}-qA_{\mu})\eta[(-i\d_{\nu}-qA_{\nu})\eta]^+-m^2\eta\eta^+,
\eeq
and the tower of massive modes does not  appear.
 

\section{Outlook and Perspectives}
\label{conc}
We would like to focus on two  points.
The first point is the criticism against the geodesic approach.  We have stressed how this approach leads to a misidentification of the rest mass and we have seen how  this ambiguity causes the $q/m$ bound and the tower of huge massive modes. Our opinion  is that the assumptions we implicitly made in GR in this approach , are no longer valid in the 5D space. The reason is the violation of the 5D PE and the impossibility to localize a particle within the extra dimension. Indeed, eq. (\ref{act5}) loses its link with the dynamics of the particle: the particle motion in 5D is not provided by the  5D geodesic trajectory. At the same time,  equation (\ref{5ddisp}) lacks of physical meaning.
The second point is a proposal for a new approach, based on a the attempt to find a rigorous definition of the 5D test particle and its rest mass.
Let us consider separately scenarios concerning the scalar  field $\phi$.
If we take $\phi=1$ from the beginning, the revised approach provides exactly the electrodynamics scenario for a reduced 4D test particle.
Mass and charge are defined in terms of independent degrees of freedom, provided by the 5D matter tensor , and no bound arises. Moreover, calculating the dispersion relation arising from this new equation, and analyzing the corresponding Klein-Gordon dynamics we see that the tower of huge massive modes is removed. The key point is now that  we do not have any link between the fifth component of the momentum and the rest mass.
If we admit a generic scalar field $\phi$ the same kind of conclusions holds but now new features appear: we deal also with a scalar field $A$, which plays a role in the coupling with matter and , more important, we now have a variable mass. This is not surprising, because the PE does not hold. Anyway, there exist possible scenarios depending on $\phi$ and $A$ where the mass is restored as a constant or , at least, we recover the free falling universality. In our opinion this new approach, even if not yet definitive, offers a solution to the matter problem in the framework of a compactified KK scenario with a small extra dimension.
Thus, from a theoretical point of vie,  it enforces the physical meaning of KK theories and deserves a detailed investigation, also in view of its multidimensional extension. A promising perspective appears the search of correct currents associated with gauge symmetries in multidimensional KK models and  the analysis of the complete cosmological solution with matter, especially when  approaching the singularity and the chaotic regime. A comparison to models for dark energy in the framework of mass varying particles ( \cite{amendola}, \cite{anderson} ) is an interesting problem to be addressed too.

\section*{Acknowledgments}
Authors are grateful to Francesco Cianfrani, Orchidea Maria Lecian and CGM members for their  feedback. 
The work of V.Lacquaniti has been partially supported by a fellowship "Bando Vinci", granted from "French-Italian University". V.Lacquaniti  is grateful  to Prof. Orlando Ragnisco (RomaTre) and Prof. Pascal Chardonnet (LAPTH) for interesting discussions.

\end{document}